\documentstyle[aps,twocolumn,epsfig]{revtex}
\begin{document}
\twocolumn[\hsize\textwidth\columnwidth\hsize\csname @twocolumnfalse\endcsname
\draft

\title{Two-Dimensional Wigner Crystal in Anisotropic Semiconductor}

\author{Xin Wan and R. N. Bhatt}
\address{Department of Electrical Engineering, Princeton University,
        Princeton, NJ 08544}
\date{\today}
\maketitle
\begin{abstract}
We investigate the effect of mass anisotropy on the Wigner
crystallization transition in a two-dimensional (2D) electron gas. 
The static and dynamical properties of a 2D Wigner crystal have been
calculated for arbitrary 2D Bravais lattices in the presence of
anisotropic mass, as may be obtainable in Si MOSFETs with (110) surface. 
By studying the stability of all possible lattices, we find significant
change in the crystal structure and melting density of the electron
lattice with the lowest ground state energy. \\ 

\end{abstract}

%\pacs{PACS numbers: 71.30.+h}
]

One of the most unexpected discoveries in the 2D electron systems in
semiconductor structures is the possibility of a metal-insulator
transition (MIT) first suggested by experiments of Kravchenko {\it et
al.} in high mobility Si metal-oxide-semiconductor field-effect
transistors (MOSFETs) in zero magnetic field\cite{kravchenko}. 
Later, a number of groups have reported possible MITs in 2D electron or
hole systems in several different semiconductor heterostructures, as well as 
in MOSFETs\cite{popovic,pudalov,coleridge,papadakis,hanein,simmons,yoon}. 
However, a genuine MIT in two dimensions is in contradiction to the scaling
theory of localization, which predicts that in the absence of
electron-electron (or hole-hole) interactions no true metallic behavior
is possible in two dimensions with pure potential scattering\cite{aalr}. 
In fact, no fundamental principle requires that such a scaling argument holds
in the presence of strong interactions, which can be measured by the
ratio of the Coulomb energy to Fermi energy, expressed by the
dimensionless parameter $r_s = m^* e^2 / \hbar^2 \epsilon \sqrt{n
\pi}$, and there have been a number of suggestions that this is the
case\cite{finkelstein,dobrosavljevic,chakravarty}.
Theoretical models not involving MIT have also been 
proposed\cite{phillips,altshuler,sarma}. 
Of all the experimental systems, the critical $r_s$ varies from 5 in
Si$_{0.88}$Ge$_{0.12}$\cite{coleridge} up to 35 in
GaAs/AlGaAs\cite{yoon}, which clearly suggests that the Coulomb
interactions are certainly not negligible, but in fact may be the
dominating scale.

It is well known that 2D electrons crystallize into a triangular lattice
(Wigner crystalization) in the low density limit where electron-electron
interactions dominate. 
In an ideally clean 2D system, the critical $r_s$ is predicted by
Tanatar and Ceperley to be $37 \pm 5$ from quantum Monte Carlo
simulations\cite{tanatar}.  
Chui and Tanatar further found that the effect of impurities can lower
the critical $r_s$ at which solidification occurs (at least on short
length scale) to as low as $7.5$\cite{chui}.
This is in good agreement with the range of the critical $r_s$ observed
in the recent experiments. 

In this paper, we have carried out a further test on the relevance of
the Wigner crystallization phenomenon to the putative 2D MITs by
exploring the effect of mass anisotropy on Wigner crystallization, since
mass anisotropy can be presented in Si- or Ge-based semiconductor
devices. 
In particular, we have studied the ground state energy of arbitrary 2D
Bravais lattice with anisotropic mass.  
To the best of our knowledge, no result on such properties of 2D Wigner
crystal with anisotropic mass has been published so far.  

The ground state energy of a lattice in the low density limit, where
exchange process are negligible, can be written as the sum of static
Coulomb energy and vibrational zero-point energy.   
With the aid of Ewald's transformation\cite{ewald}, we can write the
static ground state energy per electron of a Bravais lattice as a sum
over its lattice sites ${\bf r}^0$: 
\begin{eqnarray}
	E_s &=& - {1 \over N} \sum_{i < j}
	{e^2 \over |{\bf r}^0_i - {\bf r}^0_j|} \nonumber \\
	&=& - {2 e^2 \over \sqrt{v}} \left [ 2 - 
	\sum_{{\bf r}^0 \neq 0} \phi_{-1/2} 
	\left ( {\pi \over v} |{\bf r}^0|^2 \right ) \right ],
\end{eqnarray}
where $\phi_n(x)$ is the Misra function\cite{misra}
\begin{equation}
	\phi_n(x) = \int_0^{\infty} dt t^n e^{-xt},
\end{equation}
and $v$ is the area of the primitive unit cell of the direct lattice. 
We assume the existence of a neutralizing background of positive charges
to avoid the divergence in the evaluation of the static energy, which is
independent of the mass anisotropy.  
In a harmonic approximation, the spectrum of lattice vibrations is
determined by  
\begin{eqnarray}
	{\cal H} &=& \sum_i \sum_{\alpha} 
	{p_{\alpha}({\bf r}^0_i)^2 \over 2m_{\alpha}} \nonumber \\
	& & + {1 \over 2} \sum_{i,j} \sum_{\alpha, \beta} 
	u_{\alpha}({\bf r}^0_i) 
\Phi_{\alpha \beta} ({\bf r}^0_i, {\bf r}^0_j) 
u_{\beta}({\bf r}^0_j),
\end{eqnarray}
where ${\bf u}({\bf r}^0) = {\bf r} - {\bf r}^0$ is the deviation from
equilibrium of the electron whose equilibrium site is ${\bf r}^0$. 
The atomic force constants $\Phi_{\alpha, \beta} ({\bf r}^0_i, {\bf
r}^0_j)$ are defined by  
\begin{equation}
\Phi_{\alpha \beta} ({\bf r}^0_i, {\bf r}^0_j) 
= \left \{ \begin{array}{ll}
	- \left . {\partial^2 v(r) \over \partial r_{\alpha} \partial r_{\beta} } \right |_{{\bf r} = {\bf r}_i^0 - {\bf r}_j^0}, & i \neq j \\
	\left . \sum_{k \neq i} {\partial^2 v(r) \over \partial r_{\alpha} \partial r_{\beta} } \right |_{{\bf r} = {\bf r}_i^0 - {\bf r}_k^0}, & i = j
	\end{array} \right .
\end{equation}
with $v(r)=e^2 / r$.
The normal mode frequencies $\omega_\lambda ({\bf q})$ are solutions of
the 2D eigenvalue problem:  
\begin{equation}
	{\bf M} \omega_\lambda^2 ({\bf q}) {\bf \epsilon}_\lambda = 
	{\bf D}({\bf q}) {\bf \epsilon}_\lambda,
\end{equation}
where the dynamical matrix ${\bf D} ({\bf q})$ is given by
\begin{equation}
	{\bf D} ({\bf q}) = \sum_{{\bf R}^0} {\bf \Phi} ({\bf R}^0) 
	e^{-i {\bf q} \cdot {\bf R}^0}.
\end{equation}
In anisotropic systems, ${\bf M}$ is a $2 \times 2$ matrix. 
The dynamical ground state energy can be obtained by integrating
zero-point energies of all modes within the Brillouin zone. 

Bonsall and Maradudin did a comprehensive study on the ground state
energies of five different crystal structures, including triangular
lattice and square lattice, for isotropic 2D electrons\cite{bonsall}. 
They found that the triangular lattice has the lowest energy in low
density limit, 
\begin{equation}
\label{triangularLattice}
	E_0 = -{2.21 \over r_s} + {1.63 \over r_s^{3/2}},
\end{equation}
in units of $Ryd = m^* e^4 / \hbar^2$.
In Eq.~\ref{triangularLattice}, the first term comes from the Coulomb
interactions between electrons sitting on the lattice sites, while the
second term is from the harmonic oscillations of electrons around their
equilibrium positions.  
They also pointed out that the transverse branch of the dispersion
relation is pure imaginary for certain directions in the square lattice,
implying a dynamical instability of this lattice.  

\begin{figure}[ht]
\centerline{
        \epsfig{figure=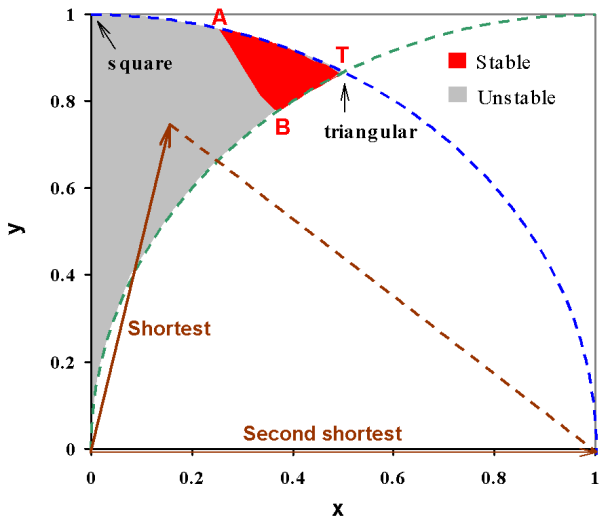,width=3.2in}
}
\caption{ 2D Bravais lattices can be mapped to points within the shaded
area. Dynamically stable lattices are mapped to only one tenth of the
irreducible area. 
\label{mapping}
} 
\end{figure}

In the presence of the mass anisotropy, the ground state may not
necessarily be one of the most symmetric lattices. 
Therefore, we have developed a systematic method to explore the ground
state energies of 2D Bravais lattices with different electron
concentrations and mass anisotropy. 
First, we map an arbitrary Bravias lattice to a point on a 2D plane by
the following scheme.   
From an arbitrary lattice site, which is chosen to be origin, one can
choose an arbitrary pair of two non-collinear lattice vectors that span
the Bravais lattice.  
If one rotates and scales the lattice so that one of the two primitive
vectors lies along x-axis normalized to unit length, the end point of
the other lattice vector represents the 2D lattice structure. 
However, a lattice can be mapped to infinite number of points due to
arbitrary choice of primitive vectors.  
One can reduce the area of points by choosing two shortest vectors from
the origin with the second shortest one placed along x-axis. 
Thus, different lattices are represented by points in the upper positive 
quadrant as shown in Fig.~\ref{mapping}.
Using reflection symmetry, however, all the points can be confined by
the y-axis and the two circles shown (the shaded area) in Fig.~\ref{mapping}. 
The three corners $(0,1)$, $\left ({1 \over 2}, {\sqrt{3} \over 2} \right )$
and $(0, 0)$ represent square lattice, triangular lattice, and
quasi-one-dimensional lattice, respectively. 
We can then sample 2D lattices by applying a rectangular mesh on the
area and calculating the ground state energy on each grid point, which
can be uniquely mapped back to a 2D Bravais lattice.  

Applying Bonsall and Maradudin's calculations on all sampled lattices
with isotropic mass, we found that most lattices have imaginary
vibrational modes, therefore are unstable.  
Under our mapping, only lattices around the triangular lattice are
stable, which occupies roughly 10\% of the reduced zone, as shown in
Fig.~\ref{mapping}.  
Using symmetry, one can actually show the triangular lattice is the
center of all stable lattices. 
Within the stable area the triangular lattice has the lowest energy for
$r_s > 25$, which is slightly below the Wigner crystallization density
$r_s = 37 \pm 5$ predicted in quantum Monte Carlo
calculations\cite{tanatar}.  
When $r_s = 25$, the lattice represented by point A (0.27, 0.95) starts
to have lower energy than the triangular lattice, whose ground state
energy is found to be 
\begin{equation}
	E_{0A} = -{2.206 \over r_s} + {1.594 \over r_s^{3/2}}.
\end{equation}
Since point A is adjacent to the unstable area, this implies that Wigner
lattice is no longer the ground state of the 2D electron system.  

\begin{figure}[ht]
\centerline{
        \epsfig{figure=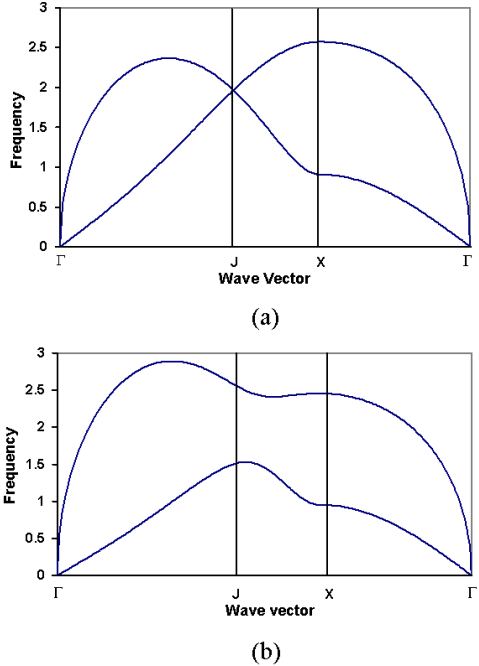,width=3.2in}
}
\caption{Dispersion curves for triangular lattice with anisotropic mass
ratio $m_l / m_t = 3$. Frequency is measured in units of $(m^* e^4 /
\hbar^3 r_s^{3/2})$.
\label{dispersion}
}
\end{figure}

We have studied different ratios between longitudinal mass $m_l$ 
and transverse mass $m_t$. 
In this paper, we present results of $m_l / m_t = 3$, which is
approximately the mass ratio in silicon (110) surface 
structures\cite{sakaki}.  
We use the geometric mean of the two mass components as the effective
mass in calculating $r_s$.  
The relative orientation of the lattice vectors with respect to the
principle axes of anisotropic mass has also been taken into
consideration.  
Figure~\ref{dispersion} compares the dispersion relation of the
anisotropic triangular lattice with that of the isotropic lattice. 
Mass anisotropy lifts the 2-fold degeneracy at J-point thus the
longitudinal branch as well. 
The ground state energy for triangular lattice fo $m_l / m_t = 3$
therefore becomes 
\begin{equation}
	E_{0T} = -{2.212 \over r_s} + {1.694 \over r_s^{3/2}}.
\end{equation}
Compared with Eq. (1), one concludes the mass anisotropy leads to an
increase in the total lattice vibrational energy. 

\begin{figure}[ht]
\centerline{
        \epsfig{figure=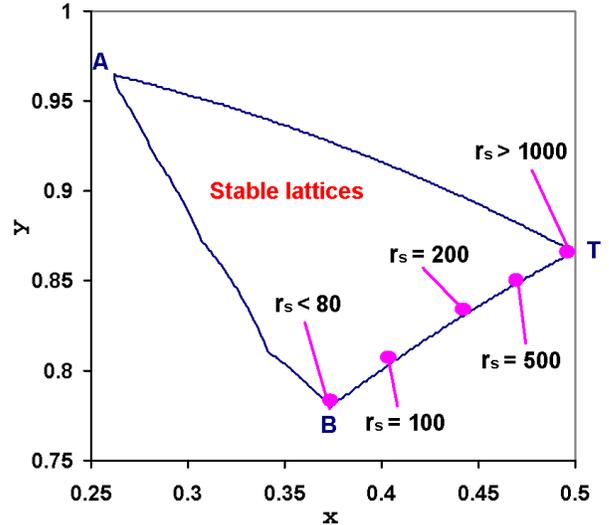,width=3.2in}
}
\caption{ The ground state move away from the triangular lattice with
decreasing $r_s$ in anisotropic system with $m_l / m_t = 3$.
\label{trend}
} 
\end{figure}

Since the electrostatic ground state energy is independent of mass
anisotropy, one expects that at low enough density, the triangular
lattice remains as the lowest energy configuration.  
However, just for $m_l / m_t = 3$ even at very low density, specially
$r_s < 1000$, we find that the lattice with the lowest ground state
energy starts to shift along one of the circular curve that confines the
reduced lattice area towards the quasi-one-dimensional lattice.  
It is understandable that a lattice can stretch along the axis of the
smaller mass to reduce the longitudinal energy of lattice vibration.  
This trend is shown in the enlarged stable area in Fig.~\ref{trend}.  

\begin{figure}[ht]
\centerline{
        \epsfig{figure=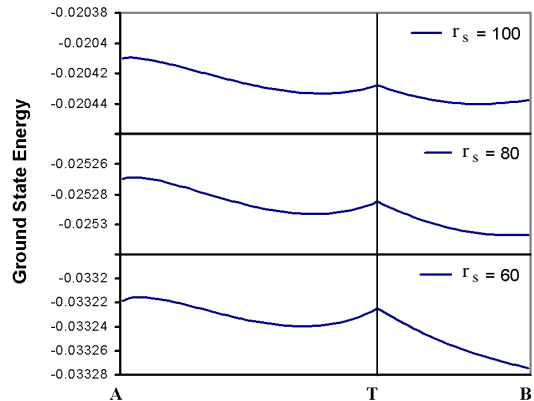,width=3.5in}
}
\caption{The ground state energy, in units of $Ryd$, along the circular
boundaries (A-T-B) for $r_s =$60, 80, and 100. 
\label{boundary}
}
\end{figure}

Fig.~\ref{boundary} shows the plots of the ground state energy for
different lattices represented by the two circular curves (A-T-B) at
different $r_s$ for $m_l / m_t = 3.0$. 
When $r_s = 100$, the ground state energy is minimized at a point along
curve TB.  
We expect 2D electrons crystallize in the lattice structure mapped to
this point at such density.  
When $r_s = 80$ the lattice with the lowest energy moves to the corner
of the stable area, B, which connects with the area of unstable
lattices.   
The comparison of the phonon dispersion curves of the triangular lattice
and the lattice mapped to point B suggests that, by crystallizing in
less symmetric lattice, the electrons gain larger energy in longitudinal
modes than the loss in transverse modes.  
When $r_s = 60$, the ground state energy decreases linearly when
approaching point B from the stable area along the circular curve TB,
which implies that no stable lattice can be formed at this density.  
The comparison of the ground state energies suggests that the Wigner
crystallization density for ideally clean 2D anisotropic electrons, such
as silicon (110) surface electrons, is much lower than that predicted
for an isotropic system. 

In this calculation we have neglected quantum mechanics of the electrons 
(i.e. electronic exchange energy). While this may have a quantitative
effect, it is unlikely to completely alter our basic conclusion that
$r_s$ of the Wigner crystallization transition increases with mass
anisotropy. This is especially true since the $r_s$ of interest ($\sim
80$) is very large, where exchange energies are smallest.

To summarize, we have found the ground state of 2D electron system has
strong dependence on the mass anisotropy. 
In a pure system, anisotropic electrons require larger spacing to form
Wigner lattice so as to stablize the long wavelength transverse modes. 
It would be interesting to see if this dramatic reduction in Wigner
crystal density can be observed in extremely clean Si (110)
MOSFET, in which our calculations estimate the critical density, in
terms of $r_s$, can be as large as 80. 

This research was supported by NSF DMR-9809483.

\end{document}